# Surface and size effect on fluctuations correlation in nanoparticles with long-range order

A.N. Morozovska [a*], E.A. Eliseev [b]

[a]V. Lashkarev Institute of Semiconductor Physics, NAS of Ukraine,

41, pr. Nauki, 03028 Kiev, Ukraine

[b]Institute for Problems of Materials Science, NAS of Ukraine,

Krjijanovskogo 3, 03142 Kiev, Ukraine

**Abstract**

Surface and size effect on the order parameter fluctuations and critical phenomena in the intensively studied *3D-confined nanosized systems with long-range order* was not considered theoretically, while the calculations for bulk samples and thick films were performed long ago. Since widely used magnetic resonance, diffraction and scattering methods collect information about both macro- and nanosystems via the structural factors, which are directly related with fluctuations correlator, *analytical expressions* for the correlation function of the order parameter fluctuations *seem extremely necessary* for quantitative analyses of the experimental data broad spectrum.

In the letter we *solve the vital problem within Landau-Ginzburg-Devonshire phenomenological approach for the particles of arbitrary shape* and consider concrete examples of the spherical and cylindrical ferroic nanoparticles. Allowing for the strong surface energy contribution, analytical expressions derived for Ornstein-Zernike correlator of the long-range order parameter fluctuations in 3D-confined system, dynamic generalized susceptibility, relaxation times and correlation radii discrete spectrum are principally different from those known for bulk system. Besides the *great importance* of the fluctuations correlation function for the analyses of scattering and magnetic resonance experimental spectra, proposed expression for fluctuations strength defines the *fundamental limit* of phenomenological theory applicability for 3D-confined nanosystems.

From the early 70-th numerous successful theoretical studies of size effects in confined systems were performed [1, 2, 3]. It was proved that phenomenological theory adequately describes size-driven phase transition in terms of the order parameter temperature and size behavior for thin films in different conditions [4, 5, 6, 7] and nanoparticles of different shape [8, 9, 10, 11, 12, 13]. These studies are complementary to the ab-initio calculations (see e.g. [14, 15]), since the usage ranges of these two approaches seem different. However in many cases the results of phenomenological models considering the long-range order parameter are shown to reproduce the results of first principles calculations [16, 17, 18].

---

[*] morozo@i.com.ua

In contrast to the size-driven phase transitions, very little attention was paid to theoretical study of the order parameter fluctuations and their correlations in 3D-confined micro- and nanosized systems, while the situation in bulk samples and thick films was considered long ago in application to inelastic light scattering from surfaces [19, 20, 21]. The absence of the appropriate theory is a vital problem, since fluctuations and correlation effects play the crucial role especially near the point of size-driven phase transition in all ferroic particles, at that their physical properties can be described in terms of the long-range order parameter behavior [22, 23]. Since modern and widely used diffraction and scattering experimental methods directly measure ***dynamical structural factors***, which are determined by fluctuation correlations in both macro- and nanosystems, ***analytical expressions*** for the order parameter ***fluctuations*** and their ***correlation function spectra*** seem extremely necessary for quantitative analyses of light scattering data [24,] and inelastic neutron scattering spectra [25, 26], as well as for the spectral data collected by EPR [27, 28, 29], NMR [30, 31], XRD [32, 33], electro- and magneto- capacitance measurements [34, 35, 36].

Besides the great practical importance of fluctuations correlation functions for quantitative analyses of scattering and magnetic resonance experimental results, expression for fluctuations strength define the ***fundamental limit*** of phenomenological theory applicability for concrete 3D-confined (meso, micro or nano) system. Actually, analytical expression of the mean square fluctuation $\langle \delta\eta^2 \rangle$ should be compared with the mean square of the order parameter $\overline{\eta}^2$. Then the inequality $\langle \delta\eta^2(\mathbf{R},T) \rangle < \overline{\eta}^2(\mathbf{R},T)$ defines the temperature range $T$ and sizes $\mathbf{R}$ necessary for phenomenological theory applicability in accordance with conventional Ginzburg criteria [37]. Below we study the problem for the particles of arbitrary shape and consider concrete examples of spherical and cylindrical ferroic (ferroelectric and/or ferromagnetic) nanoparticles.

The probability distribution of the ***fluctuation*** $\delta\eta$ of the ***one-component order parameter*** $\eta$ (spontaneous polarization or magnetization component, rotation angle, strain value, etc) is $W[\delta\eta] \sim \exp(-\delta F[\delta\eta]/k_B T)$, where $\delta F = F[\eta + \delta\eta] - F[\eta]$ is the deviation of the free energy $F$. For correct phenomenological description of any ***confined*** and especially ***nanosized*** system the ***surface energy*** should be considered, at that its contribution increases with the system size decrease. Including the surface energy term $F_S$, Landau-Ginzburg-Devonshire free energy $F$ depends on the order parameter $\eta$ as:

$$F[\eta] = \int_S d^2r \frac{\alpha_S}{2}\eta^2 + \int_V d^3r \left( \frac{\alpha(T)}{2}\eta^2 + \frac{\beta}{4}\eta^4 + \frac{g}{2}(\nabla\eta)^2 - E_e\eta - \frac{\eta}{2}\hat{E}_d[\eta] \right). \qquad (1)$$

For the sake of simplicity the surface energy coefficient $\alpha_S$ is regarded positive, isotropic and weekly temperature dependent, thus higher terms can be neglected in the surface energy expansion. Integration is performed over the system surface $S$ and volume $V$ correspondingly. Expansion coefficient $\beta > 0$ for the second order phase transitions considered hereinafter. Coefficient $\alpha(T) = \alpha_T(T - T_c)$, $T$ is the absolute temperature, $T_c$ is the phase transition temperature of bulk material. Gradient coefficient $g$ is positive. $E_e$ is the external field (e.g. electric or magnetic). $\hat{E}_d[\eta]$ stands for depolarization or demagnetization field (if any), originated from the order parameter inhomogeneity. $\hat{E}_d[\eta]$ depends on the system shape and boundary conditions, at that $\hat{E}_d[0] \equiv 0$. In general case $\hat{E}_d[\eta]$ is **linear integral operator** that exactly reduces to multiplication $\hat{E}_d[\eta] = -n_d\eta$ ($n_d$ is size-dependent **depolarization factor**) only for special case of ellipsoidal bodies with homogeneous order parameter distribution. However, for many cases effective depolarization factors may be used as a good approximation.

Minimization of the free energy (1) gives Euler-Lagrange equation with boundary conditions for the equilibrium spatial distribution of the order parameter $\eta$:

$$\alpha\eta(\mathbf{r}) + \beta\eta^3(\mathbf{r}) - g\Delta\eta(\mathbf{r}) - \hat{E}_d[\eta] = E_e(\mathbf{r}),$$
$$\left(\alpha_S\eta + g\frac{\partial\eta}{\partial\mathbf{n}}\right)\bigg|_{\mathbf{r}\in S} = 0. \tag{2}$$

$\mathbf{n}$ is the outer normal to the surface $S$.

In harmonic approximation small spatial-temporal fluctuation $\delta\eta$ should satisfy the linear boundary problem

$$\hat{L}[\delta\eta] \equiv \left(\alpha + 3\beta\eta^2(\mathbf{r}) - g\Delta - \hat{E}_d\right)\delta\eta(\mathbf{r},t) = \delta E(\mathbf{r},t) - \Gamma\frac{\partial}{\partial t}\delta\eta(\mathbf{r},t),$$
$$\left(\alpha_S\delta\eta + g\frac{\partial\delta\eta}{\partial\mathbf{n}}\right)\bigg|_{\mathbf{r}\in S} = 0, \quad \delta\eta(\mathbf{r},0) = \delta\eta_0(\mathbf{r}). \tag{3}$$

Where $\Gamma$ is positive relaxation coefficient.

Under the physical condition of positive susceptibility, the solution of the boundary problem (3) always can be expanded on the basis of orthogonal eigen functions $f_m(\mathbf{r})$ of the operator $\hat{L}$, which obey the equation $\hat{L}[f_m(\mathbf{r})] = \lambda_m f_m(\mathbf{r})$, boundary conditions $\left(\alpha_S f_m + g\frac{\partial f_m}{\partial\mathbf{n}}\right)\bigg|_{\mathbf{r}\in S} = 0$ and can be normalized as $\int_V f_m(\mathbf{r})f_n^*(\mathbf{r})d^3r = \delta_{nm}V$. Eigen values $\lambda_m$ are positive and may be degenerated [38]. Thus, the solution of (3) is $\delta\eta(\mathbf{r},t) = \sum_m C_m(t)f_m(\mathbf{r})$, where expansion coefficients $C_m(t) = C_m^0 \exp(-\lambda_m t/\Gamma)$ for $\delta E = 0$. Then the free energy deviation $\delta F$ can be rewritten using eigen functions:

$$\delta F(t) = \int_V d^3r \left( \frac{\alpha + 3\beta\eta^2}{2}(\delta\eta)^2 + \frac{g}{2}(\nabla\delta\eta)^2 - \frac{\delta\eta}{2}E_d[\delta\eta] \right) + \int_S d^2r \frac{\alpha_S}{2}(\delta\eta)^2 \equiv$$
$$\equiv \int_V d^3r \frac{\delta\eta}{2}\left( (\alpha + 3\beta\eta^2)\delta\eta - g\Delta\delta\eta - E_d[\delta\eta] \right) = \frac{V}{2}\sum_m \lambda_m C_m^2(t) \quad (4)$$

It is clear from Eq.(4) that fluctuations $C_m^0$ are statistically independent, so that the cross-probability $W[\delta\eta] = \prod_m w(C_m, \lambda_m)$, where the single probabilities $w(C_m, \lambda_m) = \frac{\lambda_m}{\Gamma}\sqrt{\frac{\lambda_m V}{2\pi k_B T}}\exp\left(-C_m^2 \frac{\lambda_m V}{2k_B T}\right)$ are independent and normalized. Corresponding correlator $\overline{\langle C_m(0)C_n(t)\rangle}$ was obtained after statistical and temporal averaging with probability $W$:

$$\overline{\langle C_m(0)C_n(t)\rangle} = \delta_{mn}\int_0^\infty dt \exp\left(-\frac{\lambda_m}{\Gamma}t\right)\int_{-\infty}^\infty dc \cdot w(c, \lambda_m) \cdot c^2 = \frac{k_B T}{V}\frac{\delta_{mn}}{\lambda_m}. \quad (5)$$

Eigen values $\lambda_m(\mathbf{R}, \alpha_S, g)$ depend on the system sizes $\mathbf{R}$, surface energy coefficient $\alpha_S$ and gradient coefficient $g$ via the boundary conditions. Therefore Eq.(5), derived for 3D confined systems with the strong surface influence, significantly differs from the conventional expression $\langle |C_k^2|\rangle = \frac{(k_B T/V)}{\alpha(T) + 3\beta\overline{\eta}^2 + gk^2}$ derived by Ornstein and Zernike [39] for the spatial fluctuations in the bulk system with eigen functions $f_k(\mathbf{r}) \sim \exp(i\mathbf{kr})$ and continuous eigen values $k = |\mathbf{k}|$. At given temperature $T$ the critical sizes $\mathbf{R}_{cr}(T)$ of the size-driven phase-transition (i.e. $\overline{\eta}(\mathbf{R}_{cr}, T) \to 0$) can be determined from the condition of zero minimal eigen value: $\lambda_{\min}(\mathbf{R}_{cr}, \alpha_S, g, \alpha(T)) = 0$ (compare with the condition $\alpha(T) = 0$ for bulk). Actually Eq.(5) is the expression for ***Ornstein-Zernike correlator*** of the long-range order parameter fluctuations in arbitrary ***3D-confined system with surface energy contribution***.

The spatial-temporal fluctuation $\delta\eta(\mathbf{r}, t) = \int_0^t d\tau \int_V G(\mathbf{r}, \mathbf{r}', t - \tau)\delta E(\mathbf{r}', \tau)d^3r'$ is determined by the ***Green function*** $G(\mathbf{r}, \mathbf{r}', t)$ typically called ***dynamic generalized susceptibility.*** For the 3D-confined system we derived expressions for $G(\mathbf{r}, \mathbf{r}', t)$:

$$G(\mathbf{r}, \mathbf{r}', t) = \frac{1}{V}\sum_m \frac{f_m(\mathbf{r})f_m^*(\mathbf{r}')}{\Gamma}\exp\left(-\frac{\lambda_m}{\Gamma}t\right). \quad (6)$$

In general case $G(\mathbf{r}, \mathbf{r}', t)$ does not depend on the difference $(\mathbf{r} - \mathbf{r}')$. This is the main distinction between expression (6) and generalized susceptibility of bulk material, $G_b(\mathbf{r} - \mathbf{r}', t) \sim \exp(-|\mathbf{r} - \mathbf{r}'|/r_B^c - t/\tau_B)$. At that bulk correlation radius $r_B^c$ and bulk relaxation time $\tau_B$ differ from the ***infinite series of size-dependent*** characteristic radii $r_m = \sqrt{g/\lambda_m}$ and relaxation times $\tau_m = \Gamma/\lambda_m$. However the critical slowing down appears for $\tau_1(\lambda_{min})$ at critical size $\mathbf{R}_{cr}(T)$ since $\tau_1(\mathbf{R}_{cr}, \alpha_S, g) \to \infty$ (see e.g. Figs.1a,c). Below we also demonstrate that ***nanoparticle correlation length*** could be estimated as $r_N^c(\mathbf{R},T) \approx \sqrt{g/\lambda_{min}}$ and diverges at $\mathbf{R}_{cr}(T)$ (see e.g. Fig.2a).

In accordance with ergodic principle and convolution theorem, the frequency spectrum of generalized susceptibility coincides with statistical averaging of the fluctuations $\delta\eta(\mathbf{r},t) = \sum_m C_m(t) f_m(\mathbf{r})$ with probability $W$, namely:

$$\widetilde{G}(\mathbf{k},\mathbf{k},\omega) = \int_{-\infty}^{\infty} dt \int_V d\mathbf{r}\, e^{i(\mathbf{k}\mathbf{r}-\omega t)} \int_V d\mathbf{r'} \frac{\langle \delta\eta(\mathbf{r'}+\mathbf{r},t)\delta\eta^*(\mathbf{r'},0)\rangle}{k_B T}, \quad (7a)$$

$$\widetilde{G}(\mathbf{k},\mathbf{k'},\omega) = \frac{1}{V} \sum_m \frac{\widetilde{f}_m^V(\mathbf{k})\widetilde{f}_m^{V*}(\mathbf{k'})}{\lambda_m + i\omega\Gamma}. \quad (7b)$$

Where we introduced the spatial Fourier image $\widetilde{f}_m^V(\mathbf{k}) = \int_V f_m(\mathbf{r}) e^{i\mathbf{k}\mathbf{r}} d\mathbf{r}$ allowing for the finite particle size.

So, generalized susceptibility $G(\mathbf{r},\mathbf{r'},t)$ is the true ***correlation function*** of the order parameter fluctuations in 3D-confined system and thus Eqs.(7) is the formulation of classical limit of Callen-Welton [40] ***fluctuation-dissipation theorem*** for the ***particle of arbitrary shape***. The result allows calculating such observable quantities as optical, far-infrared, Raman and neutron scattering dynamic structural factors, dynamic susceptibility frequency spectrum reflecting the system response to external excitation, as well as mean squire fluctuation of the order parameter. Corresponding expressions are summarized in Tab. 1 and compared with the ones for bulk system.

**Table 1.** Observable quantities related with correlation effects in particles and bulk material

| Physical quantity relation with generalized susceptibility $G$ from Eqs.(6) | Expressions for system with scalar long-range order parameter $\eta$ | | Experimental methods for direct determination |
|---|---|---|---|
| | Particle with finite sizes $\mathbf{R}$, volume $V$ | Bulk system (volume $V \to \infty$) | |
| Correlation function of the order parameter fluctuations $\delta\eta$ $\widetilde{G}(\mathbf{k},\mathbf{k'},\omega) = \int_{-\infty}^{\infty} dt \int_V d\mathbf{r} \int_V d\mathbf{r'}\, e^{i(\mathbf{k}\mathbf{r}-\mathbf{k'r'}-\omega t)} \times G(\mathbf{r},\mathbf{r'},t)$ | $\sum_m \frac{\widetilde{f}_m^V(\mathbf{k})\widetilde{f}_m^{V*}(\mathbf{k'})}{(\lambda_m + i\omega\Gamma)\cdot V}$ discrete eigen values $\lambda_m(\mathbf{R},\alpha_S,g)$ | $\frac{\delta(\mathbf{k'}-\mathbf{k})}{\alpha + 3\beta\overline{\eta}^2 + gk^2 + i\omega\Gamma}$ continuous eigen values $k$, $\delta$ is Dirac delta-function | Dynamical structural factor of optical, Raman and neutron scattering $I \sim \widetilde{G}(\mathbf{k},\mathbf{k},\omega)$ The second moment of NMR and EPR spectral line $\langle \delta H^2 \rangle \sim \widetilde{G}(\mathbf{k},\mathbf{k},\omega)$ |
| Correlation radius $r^c$ of the order parameter fluctuations | $r_N^c(\mathbf{R}) \approx \sqrt{\frac{g}{\lambda_{min}}}$ | $r_B^c = \sqrt{\frac{g}{\alpha + 3\beta\overline{\eta}^2}}$ | |
| Relaxation time(s) $\tau$ of the system response to external excitation | $\tau_m(\mathbf{R}) = \frac{\Gamma}{\lambda_m}$ | $\tau_B = \frac{\Gamma}{\alpha + 3\beta\overline{\eta}^2}$ | Phonon spectra, optical and dielectric spectra, relaxation spectra |
| Dynamic susceptibility, $\chi = (\partial \eta/\partial E_e)$, spectral density $\widetilde{\chi}(\mathbf{k},\omega) = \int\limits_{-\infty}^{\infty} dt \int\limits_{V} d\mathbf{r} e^{i(\mathbf{kr}-\omega t)} \int\limits_{V} d\mathbf{r}' G(\mathbf{r},\mathbf{r}',t)$ | $\dfrac{1}{V}\sum_m \dfrac{\widetilde{f}_m^V(\mathbf{k})\widetilde{f}_m^V(0)}{\lambda_m + i\omega\Gamma}$ <br> complex spectrum | $\dfrac{\delta(\mathbf{k})}{\alpha + 3\beta\overline{\eta}^2 + i\omega\Gamma}$ <br> simple Debye spectrum | Dielectric (magnetic) response spectra; electro (magneto) capacitance measurements; Integral intensity of scattering $(d\sigma/d\Omega) \sim \widetilde{\chi}(\mathbf{k},0)$ |
| Average dynamic susceptibility $\overline{\chi}(\omega) = \int\limits_{V} d\mathbf{r}\, \dfrac{\chi(\mathbf{r},\omega)}{V} = \int\limits_{V} d\mathbf{r}\int\limits_{V} d\mathbf{r}'\, \dfrac{G(\mathbf{r},\mathbf{r}',\omega)}{V}$ | $\sum_m \dfrac{\left|\widetilde{f}_m^V(0)\right|^2}{V^2(\lambda_m + i\omega\Gamma)}$ | $\dfrac{1}{\alpha + 3\beta\overline{\eta}^2 + i\omega\Gamma}$ | |
| Frequency spectrum of the order parameter fluctuation in **r**-space $\langle \delta\eta^2(\mathbf{r},\omega)\rangle = k_B T \int\limits_{-\infty}^{\infty} dt e^{-i\omega t} G(\mathbf{r},\mathbf{r},t)$ | $\sum_m \dfrac{k_B T \cdot |f_m(\mathbf{r})|^2}{V(\lambda_m + i\omega\Gamma)}$ | $\sim \dfrac{\exp(-r/r_B^c)}{(1+i\omega\tau_B)r}$ | Near field methods (optical microscopy, piezoresponce force microscopy) |
| Average fluctuation $\overline{\langle\delta\eta^2(\omega)\rangle} = \int\limits_{V} \dfrac{d\mathbf{r}}{V}\langle\delta\eta^2(\mathbf{r},\omega)\rangle$ | $\sum_m \dfrac{k_B T}{V(\lambda_m + i\omega\Gamma)}$ | $\dfrac{(k_B T/V)}{\alpha + 3\beta\overline{\eta}^2 + i\omega\Gamma} \to 0$ | Integral intensity |

Below we demonstrate how proposed Eqs.(6)-(7) should be applied for analytical calculation of the ***radial fluctuations*** correlation function [41] and related properties in ***ferroic particles.*** Analytical expressions for fluctuation correlation function details (eigen values, depolarization factors, eigen functions) in nanospheres and nanorods are summarized in Tab. 2. Size and surface effects of fluctuation correlations in nanoparticles are shown in Figs.1, 2.

It is clear from Figs. 1a,b that only the first relaxation time $\tau_{min} \equiv \tau_1(\lambda_{min})$ diverges at critical size $R_{cr}$ (the critical slowing down effect) and strongly depends on the surface energy coefficient $\alpha_S$ (compare dashed and solid curves), while the other times $\tau_m$ ($m > 1$) monotonically increase with particle radius increase and weakly depend on the surface energy. Figs.1b,d prove that the region of phenomenological approach applicability is $R > R_{cr}$ (except ultra-thin immediate vicinity of the size-driven phase transition, where $\langle\delta\eta^2\rangle > \overline{\eta}^2$).

It is clear from Fig. 2a that only the first correlation radius $r_1^c \equiv r_1^c(\lambda_{min})$ diverges at critical size $R_{cr}$ and strongly depend on the surface energy coefficient $\alpha_S$ (compare dashed and solid curves), while the other radii $r_m^c$ ($m > 1$) monotonically increase with particle radius increase and weakly depend on the surface energy. So, $r_1^c$ is the true nanoparticle correlation radius $r_N^c(\mathbf{R})$.

Fig. 2b clearly demonstrate that the frequency spectrum of dynamic susceptibility $\overline{\chi}(\omega) \sim \widetilde{G}(k=0,\omega)$ is governed by the surface energy expansion coefficients (compare dotted, dashed and solid curves). In contrast to bulk system the frequency spectrum can be non-Debye due to infinite series of relaxation times originated from discrete eigen values spectrum that appears from the surface energy contribution. The surface energy contribution also leads to dramatic changes in the spatial spectrum of the structural factor $I(k) \sim \widetilde{G}(k,k,\omega=0)$ (compare Fig. 2c for small value $\alpha_S = 0.1$ m$^2$/F and Fig. 2d for the higher value $\alpha_S = 1$ m$^2$/F). These effects are most pronounced for nanoparticles sizes near the critical one as anticipated.

**Table 2.** Correlation function details for spherical and cylindrical particles

| | **Sphere of radius $R$** | **Rod of radius $R$ and length $h$** |
|---|---|---|
| **Order parameter** | $\eta(r)$ is radially distributed, where $r = \sqrt{x^2 + y^2 + z^2}$ is radius | $\eta(\rho,z)$ is directed along the rod axis $z$, where $\rho = \sqrt{x^2 + y^2}$ is polar radius |
| **Depolarization or demagnetization factor** | Factor $n_d(R) \approx \zeta/(3\varepsilon_0 \varepsilon_{33}^b)$, factor $0 < \zeta < 1$ is determined by the ambient screening, $1 < \varepsilon_{33}^b < 10$ is background permittivity | Factor $n_d(R,h) \approx \dfrac{\zeta}{\varepsilon_0 \varepsilon_{33}^b \left(1 + (h/2R)^2\right)}$, the expression has high accuracy for sizes $R \ll h$ |
| **Eigen values $\lambda_m$** | $\lambda_m = \alpha(T) + 3\beta \bar{\eta}^2 + g\dfrac{s_m^2}{R^2} + n_d(R)$ <br> numbers $m = 1, 2, 3, \ldots$ | $\lambda_{m,n} = \alpha(T) + 3\beta \bar{\eta}^2 + g\left(\dfrac{q_n^2}{h^2} + \dfrac{s_m^2}{R^2}\right) + n_d(R,h)$ <br> numbers $m = 1, 2, 3, \ldots$; $n = 0, 1, 2, 3, \ldots$ |
| **Conditions for eigen values** | $\alpha_S \sin s_m + \dfrac{g}{R}(s_m \cos s_m - \sin s_m) = 0$, originated from the **surface energy contribution** ($\alpha_S \neq 0$) | $\alpha_S J_0(s_m) - g\dfrac{s_m}{R} J_1(s_m) = 0$, <br> $g\dfrac{q_n}{h} \tan\left(\dfrac{q_n + \pi n}{2}\right) = \alpha_S$. |
| **Eigen functions of radial fluctuations in $r$-space** | $f_m(r) = \sqrt{\dfrac{2 s_m}{3(s_m - \cos(s_m)\sin(s_m))}} \times$ <br> $\times \dfrac{R}{r} \sin\left(s_m \dfrac{r}{R}\right)$ | $f_{m,n}(\rho,z) = \dfrac{J_0(s_m \rho/R)\psi_n(z)}{\sqrt{J_0^2(s_m) + J_1^2(s_m)}}$, <br> $\psi_n(z) = \sqrt{\dfrac{2 q_n}{q_n + (-1)^n \sin(q_n)}} \cos\left(\dfrac{q_n}{h} z - \dfrac{n\pi}{2}\right)$ |
| **Eigen functions of radial fluctuations in $k$-space** | $\widetilde{f}_m^V(k) = 4\pi R^2 \sqrt{\dfrac{2 s_m}{3(s_m - \cos(s_m)\sin(s_m))}} \times$ <br> $\times \dfrac{s_m \cos(s_m)\sin(kR) - kR \cos(kR)\sin(s_m)}{k^3 R^2 - s_m^2 k}$ <br> wave vector absolute value $k = \sqrt{k_x^2 + k_y^2 + k_z^2}$ | $\widetilde{f}_{m,n}^V(k_t, k_z) = 2\pi R^2 \widetilde{\psi}_n^V(k_z) \times$ <br> $\times \dfrac{k_t R J_0(s_m) J_1(k_t R) - s_m J_1(s_m) J_0(k_t R)}{\sqrt{J_0^2(s_m) + J_1^2(s_m)}(k_t^2 R^2 - s_m^2)}$, <br> $J_{0,1}$ are Bessel functions, $k_t = \sqrt{k_x^2 + k_y^2}$, <br> $\widetilde{\psi}_n^V(k_z) = \dfrac{\sqrt{2 q_n}(i)^n h}{\sqrt{q_n + (-1)^n \sin q_n}} \times$ <br> $\times \left(\dfrac{\sin((hk_z - q_n)/2)}{hk_z - q_n} + (-1)^n \dfrac{\sin((hk_z + q_n)/2)}{hk_z + q_n}\right)$ |
| **Autocorrelation function** | $\widetilde{G}(k,\omega) = \dfrac{3}{4\pi R^3} \sum_m \dfrac{\left|\widetilde{f}_m^V(k)\right|^2}{\lambda_m + i\omega\Gamma}$ | $\widetilde{G}(k_t, k_z, \omega) = \dfrac{1}{\pi R^2 h} \sum_{m,n} \dfrac{\left|\widetilde{f}_{m,n}^V(k_t, k_z)\right|^2}{\lambda_{m,n} + i\omega\Gamma}$ |

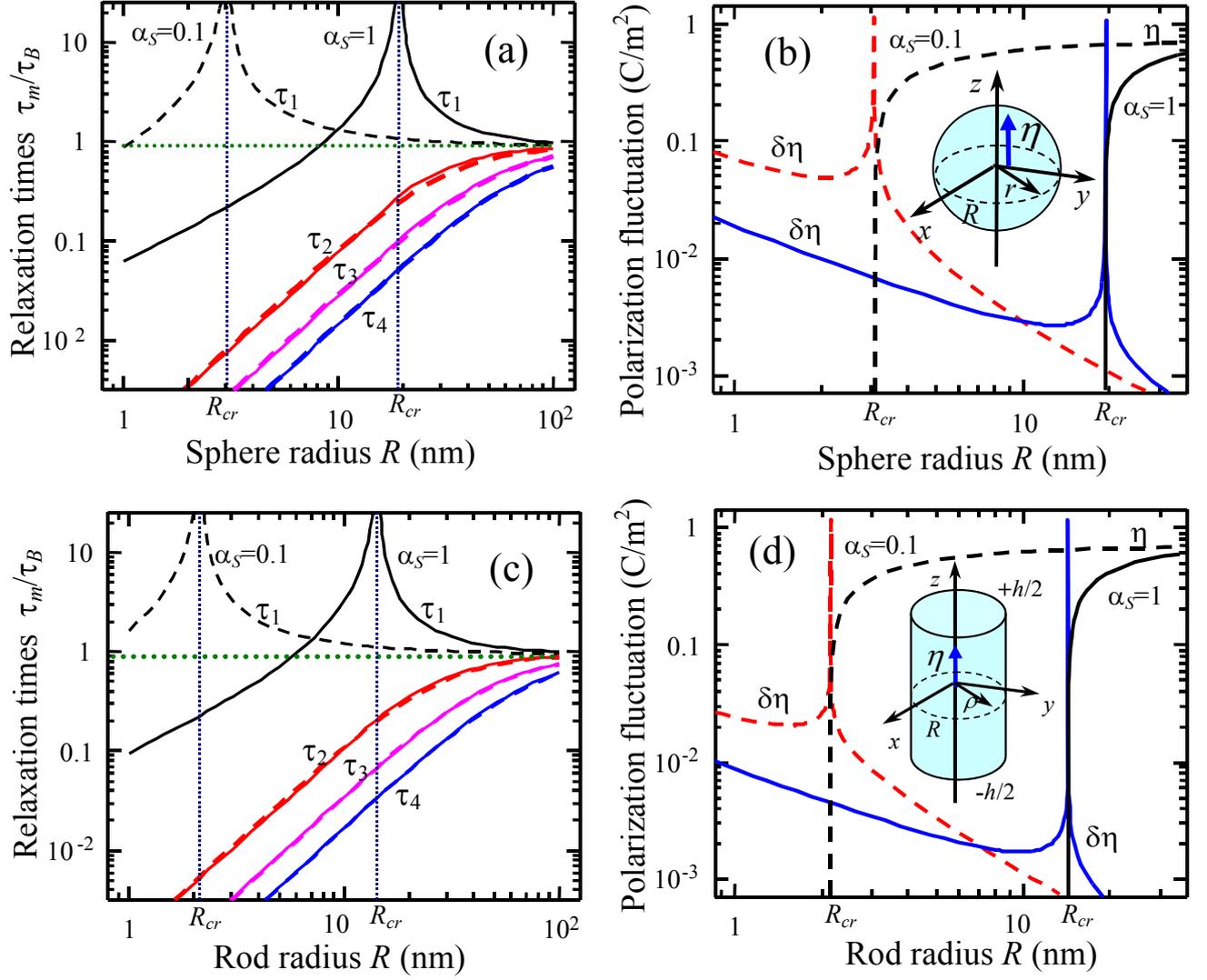

**Fig. 1**. Size dependence of relaxation times $\tau_m$ ($m$ = 1, 2, 3, 4) in nanospheres (a) and nanorods (c). The average fluctuation $\sqrt{\langle\delta\eta^2(0)\rangle}$ compared with average order parameter $\bar{\eta}$ in nanospheres (b) and nanorods (d). Material parameters $\alpha_T = 2.66 \cdot 10^5$ m/(F K), $T_C = 666$ K, $\beta = 1.90 \cdot 10^8$ m$^5$/(C$^2$F), $g = 10^{-8}$ m$^3$/F, correspond to ferroelectric PbZr$_{0.5}$Ti$_{0.5}$O$_3$, room temperature $T = 300°$K; aspect ratio $h$/R=100, small screening factor $\zeta \leq 0.01$ and different surface energy coefficients $\alpha_S$ = 0.1 and 1 m$^2$/F (solid and dashed curves).

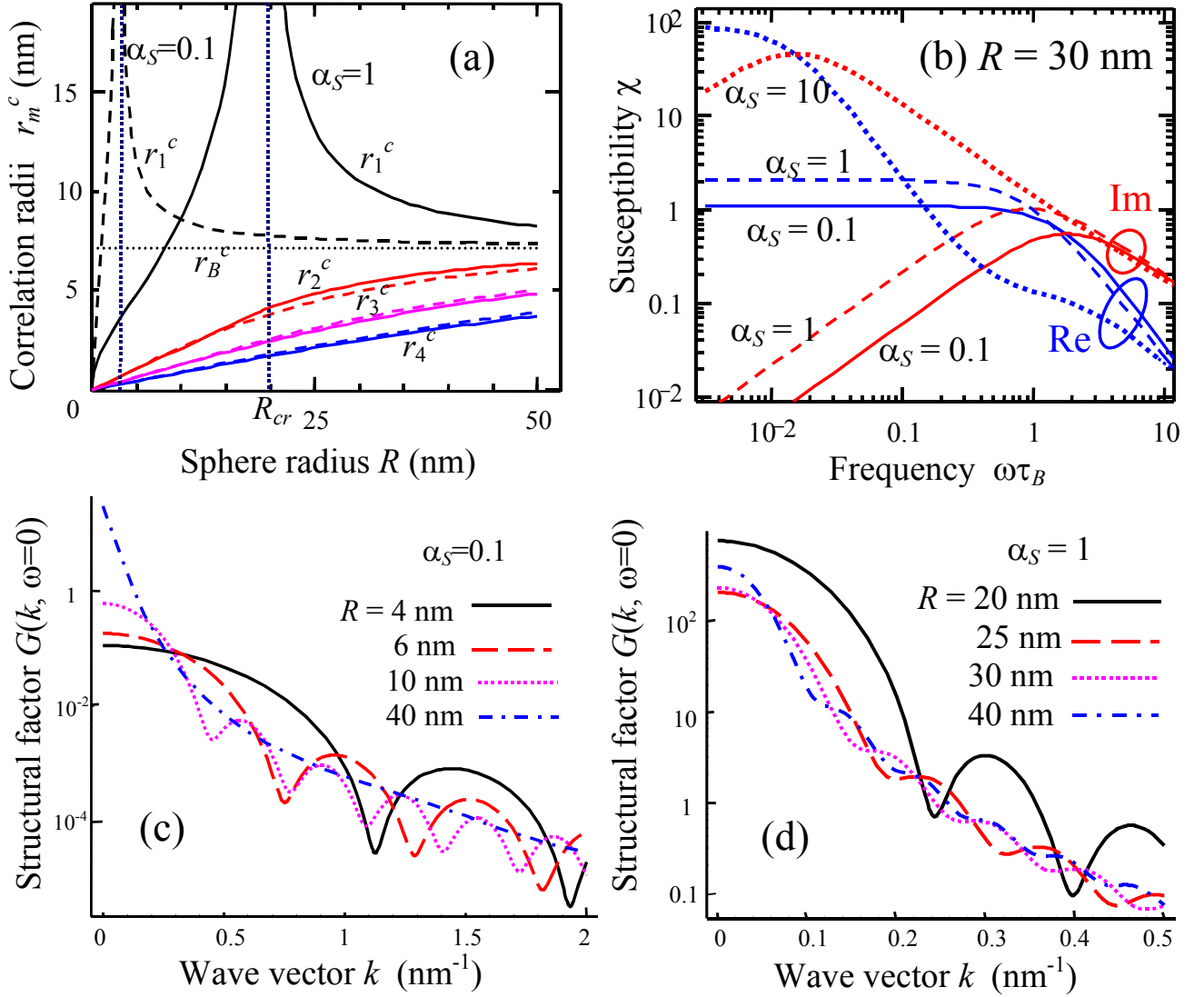

**Fig. 2**. (a) Size dependence of correlation radii $r_m$ ($m = 1, 2, 3, 4$) in nanospheres. Bulk value is marked by dotted lines. (b) Frequency spectrum of the real and imaginary parts of dynamic susceptibility $\overline{\chi}(\omega) \sim \widetilde{G}(k=0, \omega)$ for nanosphere of radius $R = 30$ nm and different surface energy coefficient $\alpha_S = 0.1, 1, 10$ m$^2$/F. (c,d) Spatial spectrum of the scattering structural factor $I(k) \sim \widetilde{G}(k, \omega = 0)$ (in relative units) for nanospheres of different radii $R = 4, 6, 10, 40$ nm and surface energy coefficient $\alpha_S = 0.1$ m$^2$/F (c) and $R = 20, 25, 30, 40$ nm and $\alpha_S = 1$ m$^2$/F (d). Other parameters are the same as in Fig. 1.

To summarize, we consider the long-range order parameter fluctuations and their correlations in nanoparticles within Landau-Ginzburg-Devonshire phenomenological theory. Obtained analytical expressions and numerical results clearly demonstrate that the fluctuations correlation function and all related physical properties (e.g. directly observable dynamical structural factors of scattering, magnetic resonance spectral line properties, dielectric and magnetic susceptibility spectra) strongly depend on the surface energy, which contribution increases with the particle sizes decrease. Since the surface and size effect on fluctuation spectra and critical phenomena in the **3D-confined systems with long-range order** were not addressed previously, our theoretical results ***provide analytical solution of the fundamental problem*** within phenomenological theory. As long as diffractional, scattering and magnetic resonant experimental methods collect information about examined systems via the structural factors directly related with fluctuation correlations, derived *analytical expressions* for the order parameter fluctuations and correlation function spectra ***could be of great importance*** for quantitative analyses of the wide range of experimental results in ***nanosized ferroics.***

## Appendix A

Derivation of Eq.(7a)

$$\int_{-\infty}^{\infty} dt \int_V d\mathbf{r}\, e^{i(\mathbf{kr}-\omega t)} \int_V d\mathbf{r}' \frac{\langle \delta\eta(\mathbf{r}'+\mathbf{r},t)\delta\eta^*(\mathbf{r}',0)\rangle}{k_B T} = \sum_{m,n} \int_V d\mathbf{r}\, e^{i\mathbf{kr}} \int_V d\mathbf{r}'\, f_m(\mathbf{r}'+\mathbf{r}) f_m^*(\mathbf{r}') \int_{-\infty}^{\infty} dt \frac{\langle C_m(t) C_n^*(0)\rangle}{k_B T} e^{-i\omega t}$$

$$= \sum_{m,n} \tilde{f}_m^V(\mathbf{k}) \tilde{f}_n^V(-\mathbf{k}) \int_{-\infty}^{\infty} dt \frac{\langle C_m(t) C_n^*(0)\rangle}{k_B T} e^{-i\omega t} = \frac{1}{V}\sum_m \frac{\left|\tilde{f}_m^V(\mathbf{k})\right|^2}{\lambda_m + i\omega\Gamma} \equiv \widetilde{G}(\mathbf{k},\mathbf{k},\omega).$$

*1) Correlation function in nanospheres* of radius *R*. The order parameter (spontaneous polarization or magnetization vector) is directed along axis *z*. Nanoparticles are covered with a screening shells, which make depolarization factor $n_d(R) \approx \frac{\zeta}{3\varepsilon_0 \varepsilon_{33}^b}$ negligibly small for the screening factor $0 < \zeta \ll 1$ (non-ferroelectric or non-ferromagnetic permittivity is $\varepsilon_{33}^b$; $\varepsilon_0$ is the universal dielectric or magnetic constant). Allowing for radial symmetry Laplace operator is $\Delta = \frac{1}{r^2}\frac{\partial}{\partial r} r^2 \frac{\partial}{\partial r}$. Normalized orthogonal eigen functions acquire the form of $f_m(r) = \sqrt{\frac{2 s_m}{3(s_m - \cos(s_m)\sin(s_m))}} \frac{R}{r} \sin\left(s_m \frac{r}{R}\right)$, number *m* = 1, 2, 3

Corresponding eigen values

$$\lambda_m(T,R) = \alpha(T) + 3\beta \overline{\eta}^2 + g \frac{s_m^2}{R^2} + n_d(R). \tag{A.1}$$

Boundary conditions at $\rho = R$ gives equation for $s_m$ determination:

$$\alpha_S \sin(s_m) + \frac{g}{R}\left(s_m \cos(s_m) - \sin(s_m)\right) = 0. \tag{A.2}$$

In high-temperature paraphase $\overline{\eta} = 0$. In the ordered low-temperature phase the averaged order parameter $\overline{\eta}$ could be found self-consistently from Eq.(2) as proposed in Refs.[10, 13]. For the minimal eigen value we derived approximation $\lambda_{0,1} \approx 2|\alpha(T) + g(s_1^2/R^2) + n_d(R)|$.

Fourier image of the eigen functions $f_m(r)$ is

$$\tilde{f}_m^V(k) = 4\pi R^2 \sqrt{\frac{2s_m}{3(s_m - \cos(s_m)\sin(s_m))}} \frac{s_m \cos(s_m)\sin(kR) - kR\cos(kR)\sin(s_m)}{k^3 R^2 - s_m^2 k}, \quad (A.3)$$

Where $k = \sqrt{k_x^2 + k_y^2 + k_z^2}$. Autocorrelation function (7b) acquires the form

$$\tilde{G}(k,\omega) = \frac{3}{4\pi R^3} \sum_m \frac{\left|\tilde{f}_m^V(k)\right|^2}{\lambda_m + i\omega\Gamma}.$$

**2) Correlation function in nanorods** of radius $R$ and length $h$. The order parameter is directed along the rod axis $z$. Under the condition $h >> R$ used hereinafter, estimated depolarization factor $n_d(R,h) \approx \dfrac{\zeta}{\varepsilon_0 \varepsilon_{33}^b (1 + (h/2R)^2)}$ is negligibly small (where non-ferroelectric or non-ferromagnetic permittivity or reference state is $\varepsilon_{33}^b$; $\varepsilon_0$ is the universal dielectric or magnetic constant). Allowing for radial symmetry Laplace operator is $\Delta = \dfrac{\partial^2}{\partial z^2} + \dfrac{1}{\rho}\dfrac{\partial}{\partial \rho}\rho\dfrac{\partial}{\partial \rho}$. Normalized orthogonal eigen functions acquire the form of $f_{m,n}(\rho,z) = \dfrac{J_0(s_m \rho/R)\psi_n(z)}{\sqrt{J_0^2(s_m) + J_1^2(s_m)}}$, where z-even function $\psi_{2n}(z) = \sqrt{\dfrac{2q_{2n}}{q_{2n} + \sin(q_{2n})}} \cos\left(q_{2n}\dfrac{z}{h}\right)$ and z-odd function $\psi_{2n+1}(z) = \sqrt{\dfrac{2q_{2n+1}}{q_{2n+1} - \sin(q_{2n+1})}} \sin\left(q_{2n+1}\dfrac{z}{h}\right)$, $n = 0, 1, 2\ldots$, $m = 1, 2\ldots$, $J_{0,1}$ are Bessel functions of the zero and first orders respectively. Corresponding eigen values

$$\lambda_{n,m}(T,R,h) = \alpha(T) + 3\beta\overline{\eta}^2 + g\left(\frac{q_n^2}{h^2} + \frac{s_m^2}{R^2}\right) + n_d(R,h). \quad (A.4)$$

Boundary conditions at $\rho = R$ and $z = \pm h/2$ gives equation for $s_m$ and $q_n$ determination:

$$\alpha_S J_0(s_m) - g\frac{s_m}{R}J_1(s_m) = 0, \quad g\frac{q_{2n}}{h}\tan\left(\frac{q_{2n}}{2}\right) = \alpha_S, \quad g\frac{q_{2n+1}}{h}\cot\left(\frac{q_{2n+1}}{2}\right) = -\alpha_S. \quad (A.5)$$

In high-temperature paraphase $\overline{\eta} = 0$. In the ordered low-temperature phase the averaged order parameter $\overline{\eta}$ could be found self-consistently from Eq.(2). For the minimal eigen value we derived approximation $\lambda_{0,1} \approx 2|\alpha(T) + g(q_0^2/h^2) + g(s_1^2/R^2) + n_d(R,h)|$.

Fourier image of the eigen functions $f_{m,n}(\rho,z)$ is

$$\tilde{f}_{m,n}^{V}(k_t,k_z) = \pi R^2 \frac{k_t R J_0(s_m) J_1(k_t R) - s_m J_1(s_m) J_0(k_t R)}{\sqrt{J_0^2(s_m) + J_1^2(s_m)}(k_t^2 R^2 - s_m^2)} \tilde{\psi}_n^V(k_z), \quad (A.6a)$$

$$\tilde{\psi}_{2n}^V(k_z) = \frac{2h\sqrt{2q_{2n}}(hk_z \cos(q_{2n}/2)\sin(hk_z/2) - q_{2n}\cos(hk_z/2)\sin(q_{2n}/2))}{\sqrt{q_{2n} + \sin q_{2n}}(h^2 k_z^2 - q_{2n}^2)}, \quad (A.6b)$$

$$\tilde{\psi}_{2n+1}^V(k_z) = \frac{2ih\sqrt{2q_{2n+1}}(q_{2n+1}\cos(q_{2n+1}/2)\sin(hk_z/2) - hk_z\cos(hk_z/2)\sin(q_{2n+1}/2))}{\sqrt{q_{2n+1} - \sin q_{2n+1}}(h^2 k_z^2 - q_{2n+1}^2)}. \quad (A.7c)$$

Where $k_t = \sqrt{k_x^2 + k_y^2}$ is the transverse wave vector, $k_z$ is the longitudinal one. Autocorrelation function acquires the form $\tilde{G}(k_t, k_z, \omega) = \frac{1}{\pi R^2 h} \sum_{m,n} \frac{\left|\tilde{f}_{m,n}^V(k_t, k_z)\right|^2}{\lambda_{m,n} + i\omega\Gamma}$.

[41] Angular fluctuations can be considered similarly, but the final expressions are much more cumbersome and omitted for the sake of simplicity